\RequirePackage{fix-cm}
\documentclass{article}       
\usepackage{graphicx}
\usepackage{multirow}
\usepackage{color}

\usepackage[margin=0.5in]{geometry}
\usepackage{hyperref}
\usepackage{authblk}

\def\ra{\mbox{$\rightarrow$}}
\def\gevc2{\mbox{GeV/c$^2$}}

\begin{document}

\title{Biased rate estimates in bump-hunt searches}


\author[1,2]{William Murray}
\author[1]{Matt O'Neill}
\author[1]{Finn O'Gara}


\affil[1]{Department of Physics, University of Warwick, Coventry, United Kingdom}
\affil[2]{Particle Physics Department, Rutherford Appleton Laboratory, Didcot, United Kingdom}

\date{\today}

\maketitle

\begin{abstract}
  The cleanest way to discover a new particle is generally 
 the `bump-hunt' methodology: looking for a localised excess in a
  mass (or related) distribution. However, if the mass of the particle
  being discovered is not known the procedure of quoting the most
  significant excess seen is 'greedy', and random fluctuations from the
  background can be merged with the signal. This means that an
  observed $3\sigma$ evidence probably has the rate over-estimated  by of order 
 10\% and the mass uncertainty underestimated by more like 20\%. If
 the width of the particle being discovered is also unknown, or the
 experimental resolution uncertain, the effect grows larger.

  In the context of LHC,  the data-doubling period is now measured in years.
  If evidence for a genuine signal is obtained at the $3\sigma$ level, the true
  signal rate probably corresponds to a lower significance. The
  additional data required to produce a $5\sigma$ discovery may
  be hundreds of   fb$^{-1}$ more than might be naively expected. 

\end{abstract}

\section{Introduction}

The observation of new particles often depends upon 
a  resonant peak seen through the reconstruction of a mass
spectrum. In contrast to the popular multivariate techniques, such an
approach  allows testing of the background model at masses above
and below the putative signal. A good
example is the discovery \cite{20121,201230}  of the Higgs boson in
2012, which relied,  at least at the time of the announcement, upon the
$H\ra llll$ and $H\ra \gamma\gamma$ channels, where good mass
resolution allowed four independent and  coincident   peaks to be
observed between the two experiments. However, the mass of the new
particle is generally not known a priori and some sort of scan must be
undertaken. In the ATLAS and CMS Higgs 
procedure\cite{ATLAS:2011tau}, (also often used elsewhere) a search is made by
making a series of statistical analyses where the hypothetical
particle mass is fixed in each test but varied between them to
explore a desired range.

In general  the most extreme statistical significance found in the
explored range is reported.   In  the presence of the null
hypothesis, (i.e.  the particle considered does not exist),  this
leads to the well-known problem of the trials factor or
`look-elsewhere effect'~\cite{Gross_2010}, that, because the reported
result is selected  because it is unusual,   
the p-value is not drawn from a uniform distribution.\footnote{The use of the `local' or `global'
p-value to claim discovery is debated. The Higgs boson discovery  used local and that is also adopted here.}

However the analogous effect, the overstating of the signal size in
the presence of a genuine signal, is not generally acknowledged.
This not a question of the stopping rule: the bias encountered if the
decision to publish is based upon the significance observed. That can
be a real issue, but  the focus here 
is on
 random background fluctuations in the vicinity of the
signal, which by their nature will be more important for small signals.  If there is an excess in 
background which is at a mass close to  a real  signal   then
the mass value giving the most extreme p-value may well be 
 intermediate  between the true signal position   and the background
fluctuation, and
there is no way for the experimenter to know that this has occurred. Since
there can be fluctuations both above and below the true particle mass
there will normally be an excess in at least one direction, and so on  average the signal
rate will be overestimated. 
This effect has been previously discussed in an unpublished CDF
experiment internal note by Dorigo \&  Schmitt\cite{dorigo}  that  the current
authors only obtained very late in producing this manuscript.
In addition, if there is uncertainty about the width of the signal as
seen in the detector, which could be from unknown  natural width or
uncertain detector resolution, this too will be exploited to increase
the measured signal further. 
A corollary of this is that the observed excess, while including the
genuine signal, 
typically also contains background fluctuations, which
makes the mass estimate unreliable.

These biases  are only relevant when the signal does not dominate over
background fluctuations, but the observation of excesses of marginal
significance can attract attention. A $2.9\sigma$
excess reported by the CMS collaboration~\cite{2019320} 
in $m(\gamma\gamma)$ near 95 GeV has resulted in many
theoretical speculations,   34 of which are  collected in 
Ref.~\cite{janot_2024}, along with a useful discussion of how the
significance should evolve with more data.  That analysis
uses a scaling of significance with 
luminosity which does not account for the effect discussed in this
document  to overstate
the case against the reality of this particle --  although the author is
very likely correct in his conclusions.

This paper explores the size of these  effects in a simple example
case, with very few parameters and no systematic errors. The
methodology is discussed in Section 2, with results on rate and mass
in Sections 3 and 4 and a brief conclusion.

\section{Methodology and fit validation}

A real experiment will involve complicated models and many systematic
effects, but here a minimal study of an idealised mass spectrum is
used. The approach is simple but is described in detail to avoid misunderstandings.

The  background is taken to be uniform in mass within the
considered range  and the  signal is assumed to come from a narrow
resonance  such that the observed width
is known to the experimenter from the detector response. This is taken
here to be a  Gaussian. The rate and mass are taken as unknown,
while the expected background rate is assumed known.\footnote{Fitting
for the  background rates creates a complicating dependence of the significance  on the fit
range, but does not affect any conclusion.}
The model is fit to simulated datasets  using CERN's `root'
package~\cite{rene_brun_2019_3895860}, using the {\tt TH1::Fit()}
method which  is itself relying on `Minuit'~\cite{James:2296388}. The
fit minimises the negative log-likelihood.

The complete model is thus simply
\begin{equation}
y = a + be^{-0.5 (x-m)^2/\sigma^2},
\end{equation}
where $a$ is the background per bin,  $m$ the signal mass, $\sigma$ is
width and $b$ its height.
It is evaluated (as is root's normal behaviour) at the bin centre
only. The conversion between the fit parameters and the signal rate is
established by numerical integration using the same binning.

\begin{figure}[htb] 
 \centering 
 \includegraphics[width=0.49\textwidth]{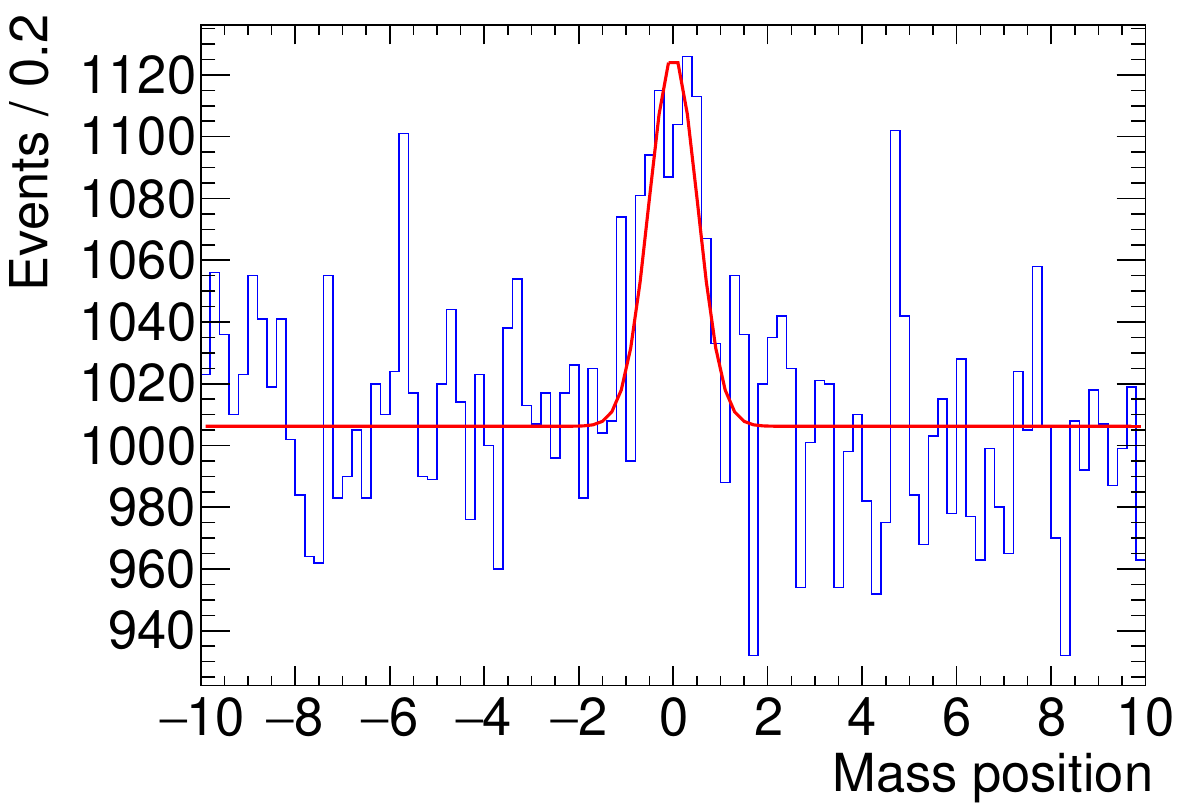}
 \includegraphics[width=0.49\textwidth]{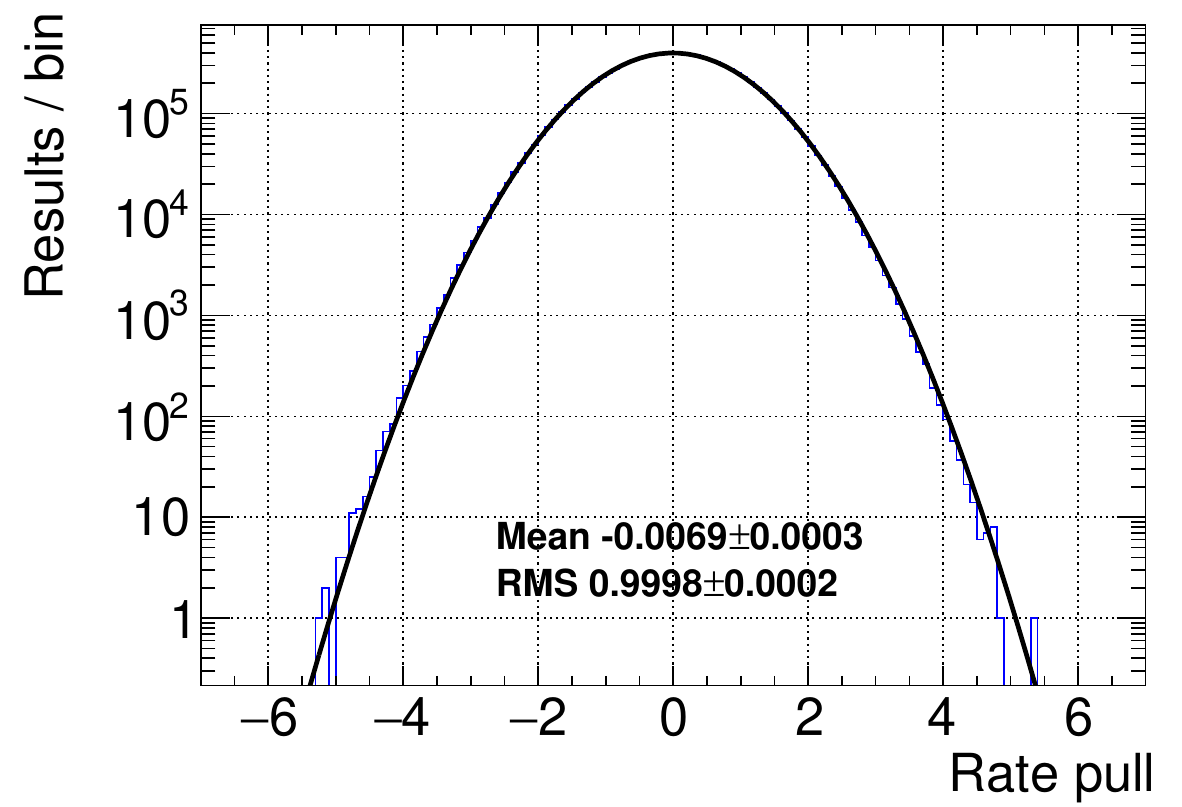}
 \vskip -0.2cm 
 \caption[]{
   Left: an example toy experiment with a large signal injected as
   mass zero.
   Right: the pull distribution of the fitted signal rate when testing
   for a signal at the origin in 1M background-only toy experiments
 } 
 \label{fig:backgroundonlydist}
 \end{figure}

Fits are performed using a background distribution covering a mass
range 2 units wider than will be scanned for a signal peak. The
Gaussian signal width is set to 0.5. This
ensures that so that there is negligible signal density  outside the fitted area.
So for a scan range of $\pm$8, a region  -10 to 10 will be fitted  in 100 bins of width 0.2. The expected background
level is normally set to 1000 events per bin, in order that the fluctuations in
content are for all practical purposes Gaussian distributed.  
An example to make this simple
set-up unambiguous is shown in Figure~\ref{fig:backgroundonlydist} (left).

The large background rate used in these tests allows for a
simplification in the procedure compared with typical LHC 
experiments.  The recommended procedure is to find  the p-value of a result by
building the distribution of the likelihood under the background only
hypothesis and then evaluating the fraction of that distribution which
gives a results equal to, or more extreme than, the one
observed. But as the per-bin fluctuations are effectively Gaussian the
errors returned by Minuit, known as `MIGRAD' errors  are used
directly, saving significant computing resource.
 To  test the accuracy of this,  10M toy experiments with no
 signal were generated and fitted for a signal in the centre, and the `pull', the ratio of the
 fitted signal size to its error was recorded.  This  cohort had a rate pull
 distribution shown in Figure~\ref{fig:backgroundonlydist} (right) to
 be 
 well described by a Gaussian with a   mean signal 
significance of  $-0.0069\pm0.0003$, showing
a small  bias, negligible for current purposes,  and an RMS
$0.9998\pm0.0002$, validating the MIGRAD error estimate.

Toy experiments were run to test signal extraction using a signal
positioned at mass zero and tested at the same mass. 
When signals drawn from a distribution with 1000 expected signal
events are  tested  the mean number of events seen is 996, with an
error below 1. This   non-closure  has not been understood,
  but it is affected by the background rate, being 997 or 995 for
  background rates of 100 or 10,000 respectively.  It is significantly
  smaller than the bias whuch is the subject of  the next section.

  Since discovery
p-values are to be extracted using the significance of the measured
signal rate, the relevant  error is not that returned from those fits, 
but the average  one  from fits to background only
experiments. To be strictly accurate the signal rate is multiplied by
the mean inverse error, although the difference between inverse mean
and mean inverse is  $1.00004$,  negligible for the current study.
  


\section{Bias in rate when scanning mass}

With the level of accuracy of the fits
when testing for a signal at the correct mass established,  the impact of
scanning for the peak position can be studied. The putative signal mass is varied
across the fitting range in steps of 0.1$\sigma$ and the fit repeated
at each point. In doing so,  0.01\% of toy experiments had a fit status
non-zero from some mass point, usually at the very edge of the range
tested. Such  experiments are entirely discarded. In the remaining
ones the point with the largest fitted 
signal rate (which by construction is also the most significant) is
selected as the point at which data will be reported.

\begin{figure}[htb] 
 \centering 
 \includegraphics[width=0.49\textwidth]{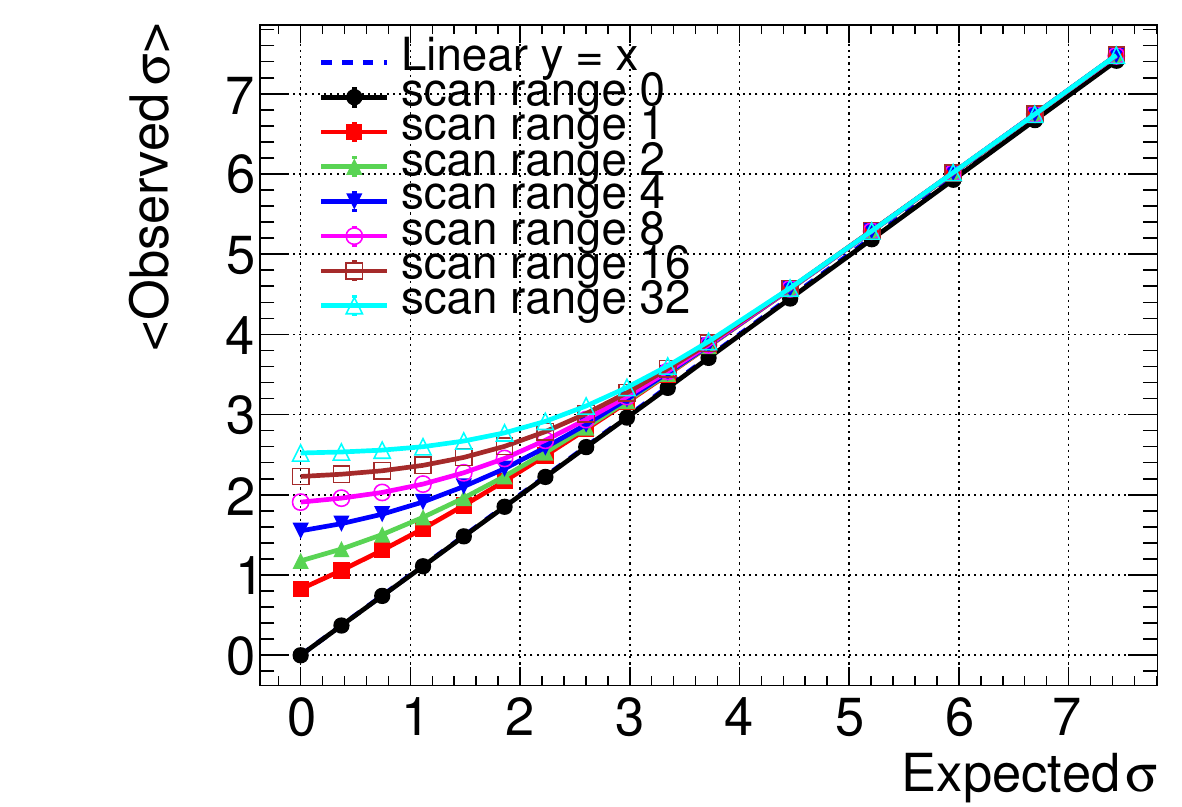}
 \includegraphics[width=0.49\textwidth]{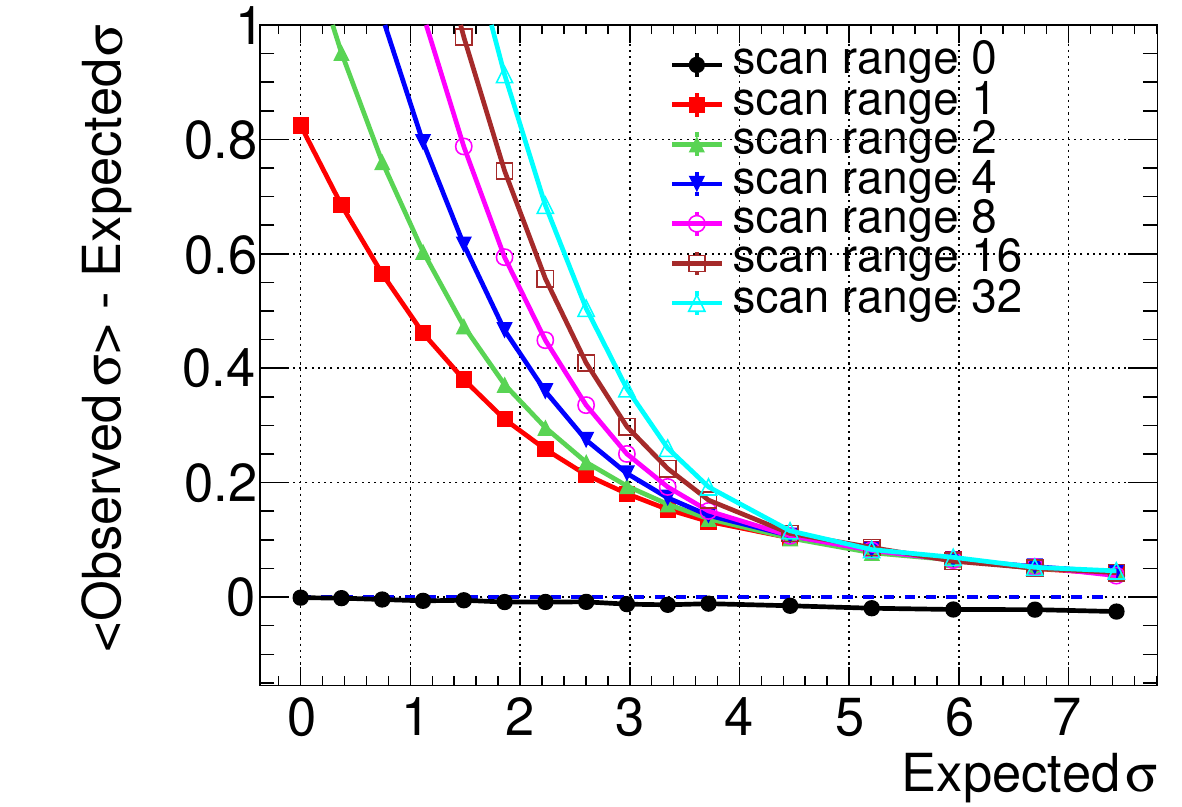}
 \vskip -0.2cm 
 \caption[]{Left: The mean signal size as a function of the expected with
   various different scanning ranges. Right: The residuals of the data
   to the straight line. The slope in the  `scan range 0' residuals
   reflects a non-closure in the study.  but the qualitative change
   when a mass range is scanned is clear.} 
 \label{fig:multi_plot} 
 \end{figure}

The dependence of expected significance on the injected signal
strength and the mass range scanned is explored in
Figure~\ref{fig:multi_plot} which shows both the mean expected
significance and the difference with that naively expected.
The signal 
 strengths range from zero  up to  700 events, which corresponds to an expected
result of just over  7$\sigma$ when testing at the correct mass
only. The scan range is from zero to $\pm$32, or 64 times the Gaussian
signal width.

 When  a range of zero
is tested, i.e. the signal is tested for only at exactly the mass at
which it was generated,   the observed significance is around 0.4\%
less than expected. This non-closure was mentioned
in the previous section.

The other data show the impact of scanning a range for the highest
excess. When the signal injected is in fact zero but the highest
signal in the range is reported then the trials
factor behaviour is recovered and the mean significance  rises with
the scan range. However, when  a signal approaching 5$\sigma$ is considered
  the bias depends largely on the {\em fact} that a range is scanned
  and has little dependence on how large it is.  This is because
  creation of such a large peak totally at random is unlikely, but the
  addition of a shoulder to the real peak is always a viable
  possibility.  For  expected signal sizes around 3 sigma or less
  the range scanned is  important because there is a 
  significant chance the maximal   peak is not connected with the signal.

\begin{figure}[htb] 
 \centering 
 \includegraphics[width=0.49\textwidth]{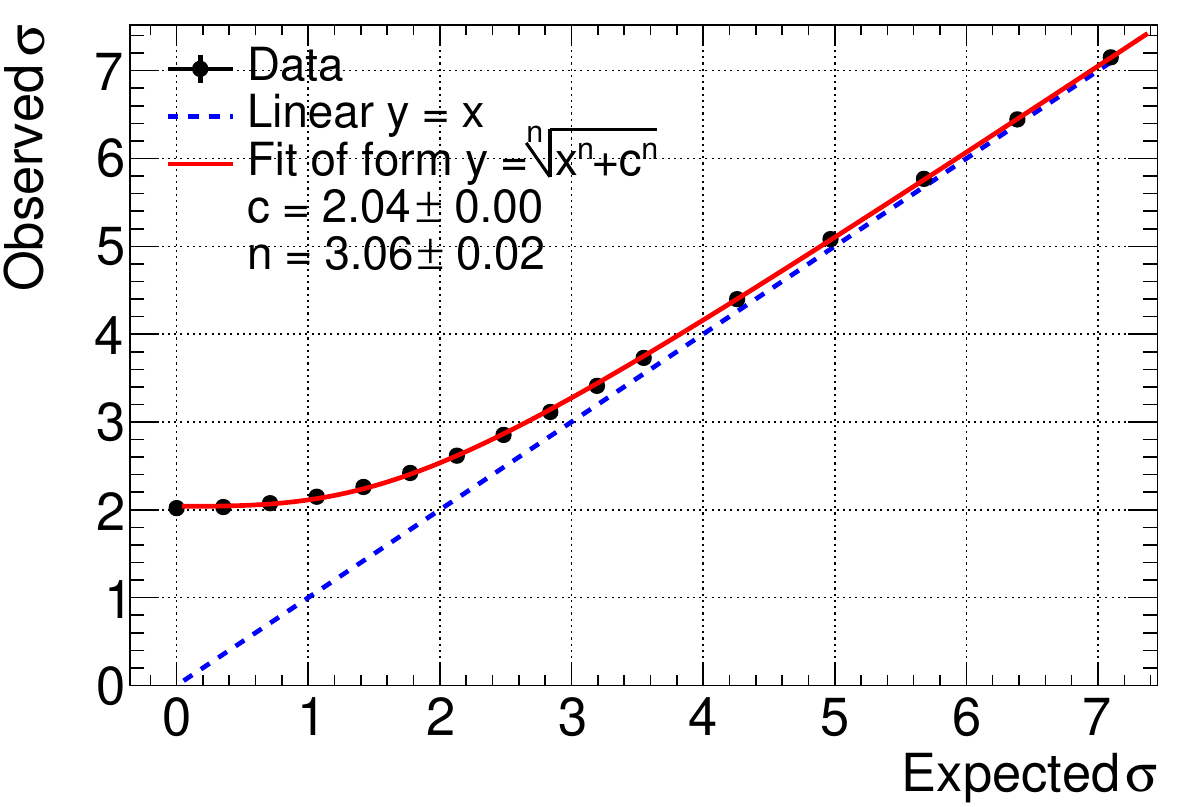}
 \includegraphics[width=0.49\textwidth]{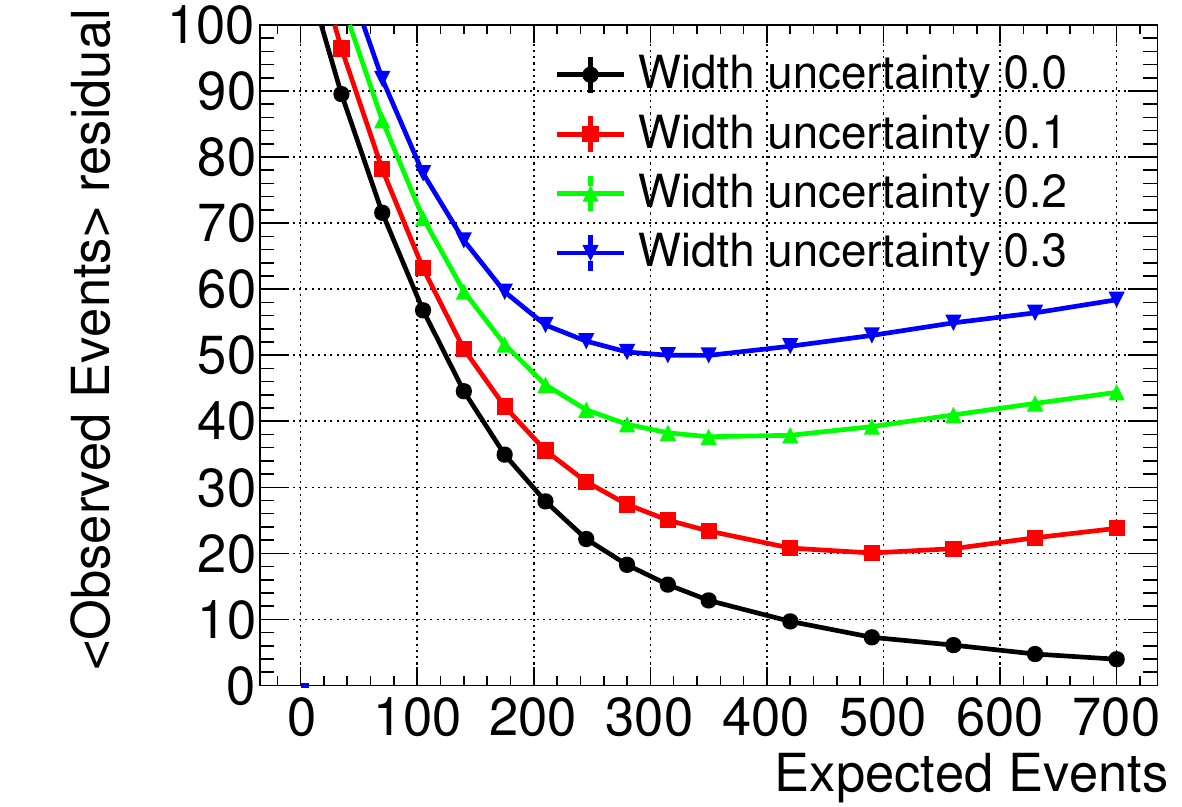}
 \vskip -0.2cm 
 \caption[]{
   Left: The mean significance found as a function of that  expected
   when the signal is tested only at    the mass at which it is
   injected.  Right: The bias seen using a mass scan range of $\pm$2 and an
 uncertainty on the signal width between  0 and  $\pm$0.3, in terms of
 the numbers of events. } 
 \label{fig:scatter_fit} 
 \end{figure}

The $H\rightarrow\gamma\gamma$ searches in
Refs.~\cite{20121,201230} had a ratio of range tested to signal width
roughly corresponding to the range of 8 tested here. 
This range is used in  Figure~\ref{fig:scatter_fit} to find an
approximate parametrisation of the size of the effect.

The observed significance should be linearly proportional to the
injected signal size for large signals, but essentially independent of
it for small ones where background fluctuations dominate. This
suggests that a plausible heuristic description might be achieved with
the  function
 \begin{equation}
y = \sqrt[n]{x^n+c^n}. \label{eq:cn}
 \end{equation}
 where $x$ is mean signal significance expected when testing only at
 the true mass, and $y$ that obtained when the most extreme result is
 reported. This  is fitted to the data in Figure~\ref{fig:scatter_fit} (left)and
provides an adequate description with $c=2.04$ (for significance)  and
$n=3.06$.
The `$c$' parameter represents the mean significance in the absence of
signal. Its size depends upon the scanned range: the most extreme
p-value in a search with no signal is inversely proportional to the
scan width.

The traditional statistical level above which data might be  presented as evidence
for a new particle is  $3\sigma$. In the case explored here, this would be
a mean result when the signal strength would provide 2.66$\sigma$
when tested only at the correct mass.
The experimenter might then overestimate the signal rate by 13\%.
and  calculate how much data would be required to reach an
expected significance of $5\sigma$ with a naive $\sqrt{\mathcal L}$
  extrapolation. This would suggest  that 
recording an additional 1.8 times the original dataset, to get a total
approximating  
$(5/3)^2$,  is sufficient. However, applying  Eqn.~\ref{eq:cn}  it is
found that 
the observed 5$\sigma$ would be expected from  an underlying $4.9\sigma$, which
requires an additional dataset 2.4 times the original. 
In other words the approach currently used  will suggest
that confirmation needs only $3/4$ of the additional data actually
required, though this depends upon the range scanned.


The background rate was initially 1000 events per bin so that Gaussian
errors could be expected. The results shown here, expressed in terms
of observed and expected significance, show no dependence upon this
number until the background level goes below about 10 expected events. At
this point the breakdown of the Gaussian approximation manifests
itself through an increase in the `$c$' parameter in
equation~\ref{eq:cn}.

The bias evaluated so far  comes from not knowing the true mass. If the
signal width is also in doubt that introduces an additional
susceptibility to fluctuations. An uncertain width might derive from
the assumed signal theory, or it might come from lack of knowledge of
the experimental resolution.
The impact is tested 
by allowing the fit to optimise  the width freely, but limited by
lower and upper bounds.   The bounds are around the central value, $0.5$,
extending by $\pm0.1$,  $\pm0.2$ or  $\pm0.3$. The scan range is set to $2$.
In the absence of signal such a fit typically returns a value for the
width of  one or other
of the limits, giving rise to a non-Gaussian, bimodal, error  distribution,  and the
calculation of significance using the mean error from such fits is not applicable.
For simplicity the results are presented in
Figure~\ref{fig:scatter_fit} (right) in terms of the signal rate
in events. The data shown in this figure in black, for no width
uncertainty, is the same as that seen in Figure~\ref{fig:multi_plot}
in green. For this curve 280  expected  events corresponds to a
$3\sigma$ expected evidence, and other points are in proportion.

The result of fitting for the width of the distribution is again to
bias the rate upwards. For  a signal of 500 events, which in the absence of
resolution uncertainty would correspond to a 5$\sigma$ excess, this
bias can be much larger than the mass uncertainty alone, potentially a
10\% effect. As a fraction, it is decreasing with signal rate, but
rather slowly.



\section{Reliability of mass fitting}

In the LHC experiments fitted masses are rarely published unless  the
evidence reaches the 5$\sigma$ discovery threshold. Nevertheless,  the
mass of less significant signals may well be  of interest to someone exploring for
example the compatibility of two channels or two different datasets.
The early Higgs boson mass 
measurements~\cite{PhysRevD.90.052004,PhysRevLett.110.081803} used
the change in -$2\times \ln \Lambda$ by 1 to estimate one sigma errors
on the mass. This is also the method used by the Minuit package~\cite{James:2296388}. The accuracy of this approach is tested here.

In order to test this in a Frequentist manner  it is necessary to  define an
ensemble of experiments. There are several ways of doing this.
It is  possible to use the true signal  rate, but this is not
applicable to the real world where the true rate is unknown.
The experimentalist knows the fitted signal size, and  might construct
an ensemble of toy simulations drawn from that hypothesis. In the light
of the previous section a correction could be made for the expected
bias in the rate. Here an ensemble is constructed with a uniform
distribution of true signal rates between zero and 8$\sigma$
expected, but the data is conditioned on the observed significance.
This reflects the experimenter who knows the significance of her
observation, but not the underlying signal cross-section.

The approach used is to generate a toy experiment, scan it for the
most extreme signal and then refit with the mass parameter
variable. For simplicity the width here is fixed to its true value, 0.5. To aid Minuit convergence, it was found helpful to fit twice,
with initial mass values as the most extreme point $\pm$0.5. The other
parameters were  initialised as returned from the  scan fit.
 In most cases the two fits converge to the same mass, if not the
 better likelihood fit was retained. A small fraction of  fits,
 0.005\%,  were  rejected  because they reported  an error below 0.005
 - far below the typical 0.1. The best
 fit mass was then found and its error taken from the MIGRAD estimate,
 and its pull, the ratio, was studied. 

In addition, the errors are evaluated by
 explicitly finding the masses at which the $\-2\times \ln \Lambda$
 increases by 1 (``method of MINOS'').
 This has most impact for toy experiments
 in the shoulders of the mass distribution, producing asymmetric
 errors typically bracketing the symmetric ones extracted from MIGRAD.
 In an approximate treatment of these non-Gaussian features
the error evaluated in the direction of the true signal
 mass is used instead of the MIGRAD error. As
 will be seen, the difference is very small.

\begin{figure}[htb] 
 \centering 
 \includegraphics[width=0.49\textwidth]{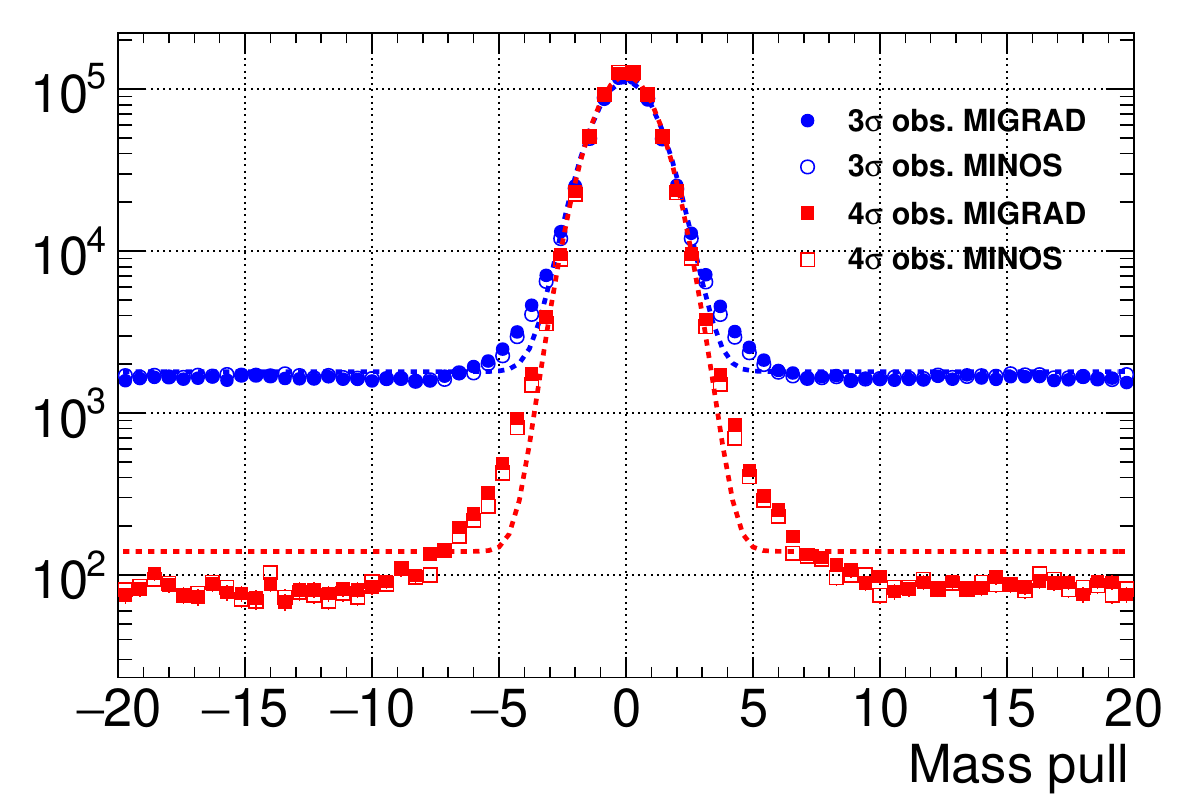}
 \includegraphics[width=0.49\textwidth]{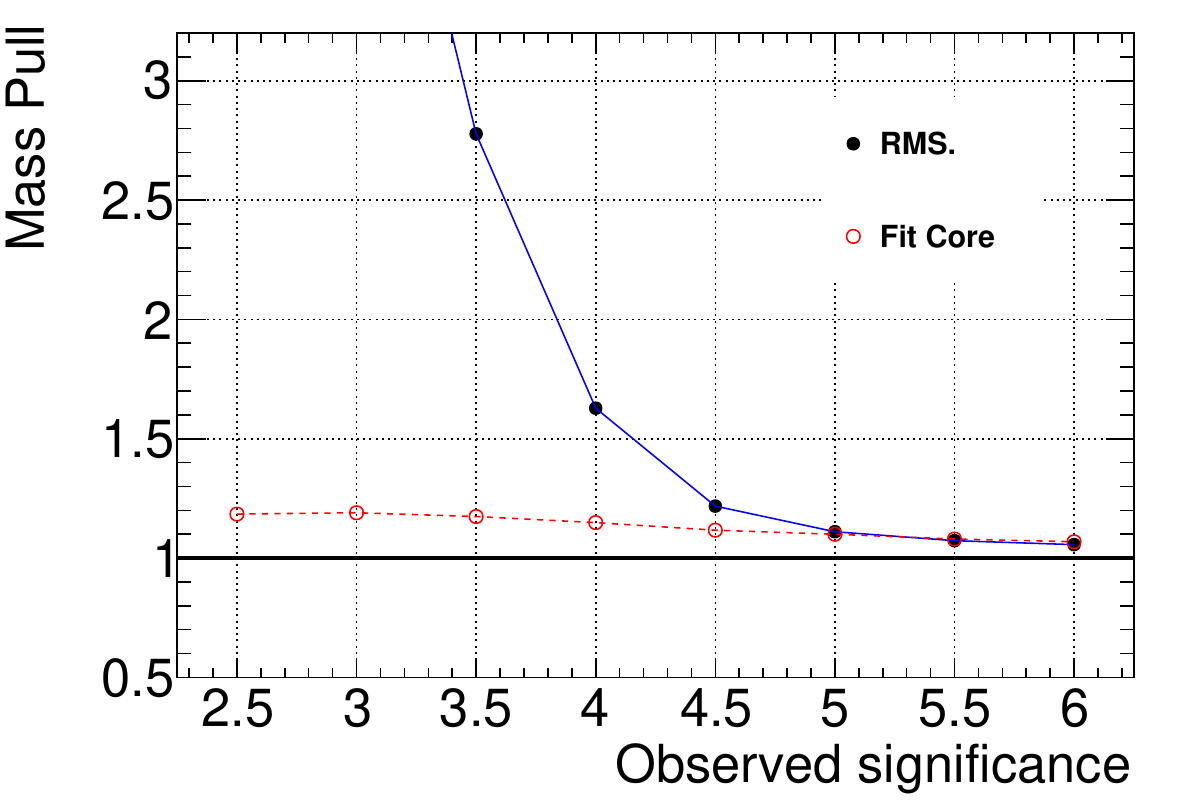}
 \vskip -0.2cm 
 \caption[]{Left: the pull in fitted mass for experiments with an observed
   significance 3$\pm0.25$ and 4$\pm$0.25 $\sigma$. Solid symbols use
   the error from MIGRAD, open for MINOS error, circles when
   the observed significance is near $3\sigma$, squares for
   $4\sigma$. The fit is to the pulls from MIGRAD using  a
   Gaussian plus a flat background to illustrate the non-Gaussian
   behaviour. Right: the RMS  and Gaussian sigma core versus the observed
 significance. } 
 \label{fig:obs34} 
 \end{figure}

 The ensemble was a set of experiments with a signal at position zero,
 with an amplitude corresponding to an expected (naive) signal significance
 uniform between 0 and 8 $\sigma$. The results  were then categorised by
 the observed significance. Figure~\ref{fig:obs34} (left) shows the pull
 distribution when the observed significance is within $0.25$ of 3 and
 4 $\sigma$.

The fit means naturally cluster around the true simulated mass, but
there is a significant,  essentially  uniform, distribution of
observations due to background fluctuations.  The RMS of the overall
distribution will depend upon the mass range searched and reflect the
well-known issue of the trials factor. More pertinent here is the
distribution of the peaking component, which is characterized by
fitting with a Gaussian. These have widths of 1.19$\pm$0.01 and
1.14$\pm$0.01 for observed significances of $3$ and 4 $\sigma$
respectively, not the unit width expected of a well-behaved error
distribution. They  also exhibit tails
beyond the Gaussian, with a significantly larger fraction having a
pull  of five or even more. The use of the MINOS
errors rather than  the MIGRAD ones  reduces the tails
noticeably at about $4\sigma$ but the difference is minor.

  Figure~\ref{fig:obs34} (right)  shows the RMS and Gaussian core
  width as a function of the observed significance. For this figure 
pulls between $\pm$10 are considered, limiting the RMS. The RMS and
fit core width deviate significantly unless the observed significance
is at least 5$\sigma$, which emphasises the non-Gaussian nature of the
error distribution. However, for significances between the `evidence'
and `discovery' thresholds even the fit core 
 uncertainty is underestimate by at least  10\%. 

The ensemble used  for these studies is tested in two ways. Firstly
the injected signal was changed so that it was flat between no signal
and 7$\sigma$ and the tests repeated. The results were compatible
within errors. More radically the ensemble was changed, so that only
experiments where both the expected and observed significances were within
3$\pm$0.25 or 4$\pm$0.25 were tested. Again the results were
compatible to within 0.01 in the width of the pull distributions. 

\section{Discussion and Conclusions}

The techniques for establishing the presence of a new particle in LHC
searches have been well tested for their performance in the presence
of the background-only hypothesis.  However, in the presence of a
genuine, small,  signal  the impact of background
fluctuations produces an upward  bias in the measured rate and
underestimated errors on the mass. This is unavoidable (though in
principle correctable of course) when the mass is unknown. 
It is made substantially worse, especially for larger signals,  if
there is also uncertainty on the resolution or width.  Such an
uncertainty might be introduced by an experimenter introducing a
'conservative' error on the detector  resolution in a misguided
attempt to avoid bias.

If the resolution is perfectly known, at the generally recognised
threshold for claiming evidence, 3$\sigma$, the rate bias can be over 10\%,
This has implications for the evolution of the significance. For
example, suppose LHC Phase I of the LHC delivers 400~fb$^{-1}$ and a three
sigma evidence in a scan range of 8. To convert this into a 5$\sigma$
discovery  then requires from Phase II  (assuming all is identical)  
not 720~fb$^{-1}$ but 960~fb$^{-1}$ . This would only increase if
there is uncertainty about the width.

Most results here are  expressed in terms of 
significance, which  has the advantage that that they are independent of
the absolute rates.
However,  a direct, proportional correspondence between
rate and significance is used  rather than estimating  $p$-values 
from toy experiments which is only applicable for linear problems and
cannot be used when the width is fitted for.
It is
therefore worth remarking that if these results are
expressed in terms of signal rate they  must, by construction, show
the same fractional bias. Signal rates are generally estimated
from the maximum  of a likelihood,  as was done here, and so
suffer this bias.

Mass estimation for particles at the threshold of discovery should be
also be taken with a grain of salt.  The distribution of pull values
shows a core that is reasonably described by a Gaussian but with a
width 10-20\% larger than  expected and significant extended tails to
$5 \sigma$. In addition there remains the possibility that the wrong
peak in the data has been taken as the signal.

The numerical importance   will depend in detail on the shape of signal and
background distributions, but the basic effect, the tendency to merge
events from  a signal at an unknown mass with any nearby fluctuation,
is universal.


\bibliographystyle{unsrt}   

\bibliography{greedy}


\end{document}